\documentclass[10pt,a4paper,twocolumn]{article}
\usepackage[utf8]{inputenc}
\usepackage[T1]{fontenc}
\usepackage[a4paper,margin=2.5cm]{geometry}
\usepackage{amsmath,amssymb,amsfonts}
\usepackage{algorithmic}
\usepackage{graphicx}
\usepackage{textcomp}
\usepackage{xcolor}
\usepackage{multirow}
\usepackage{listings}
\usepackage{fontawesome}
\usepackage{authblk}
\usepackage{float}
\usepackage{caption}
\usepackage[colorlinks=true,linkcolor=blue,citecolor=blue,urlcolor=blue]{hyperref}
\usepackage{csquotes}

\usepackage[backend=biber, style=ieee, isbn=false, sortcites, maxbibnames=6, minbibnames=1]{biblatex} 

\DefineBibliographyStrings{english}{%
  andothers = {et\addabbrvspace al\adddot}
}
\DefineBibliographyStrings{english}{
  url = {\adddot\space[Online]\adddot\space Available:}
}
\DefineBibliographyStrings{english}{
  urlseen = {Accessed:}
}
\DefineBibliographyStrings{english}{
  pages = {pp\adddot},
  page = {p\adddot}
}

\def\BibTeX{{\rm B\kern-.05em{\sc i\kern-.025em b}\kern-.08em
    T\kern-.1667em\lower.7ex\hbox{E}\kern-.125emX}}

\begin{document}

\title{\textbf{Hybrid Quantum Security for IPsec}}

\author[1]{Javier Blanco-Romero}
\author[2]{Pedro Otero Garc\'ia}
\author[1]{Daniel Sobral-Blanco}
\author[1]{Florina Almenares Mendoza}
\author[2]{Ana Fern\'andez Vilas}
\author[2]{Manuel Fern\'andez-Veiga}

\affil[1]{Department of Telematic Engineering, Universidad Carlos III de Madrid, Legan\'es, Madrid, 28911, Spain}
\affil[2]{atlanTTic Research Center (IC Lab), University of Vigo, Spain}

\renewcommand\Authands{ and }
\renewcommand\Affilfont{\itshape\small}

\maketitle

\begin{abstract}
Quantum Key Distribution (QKD) offers information-theoretic security against quantum computing threats, but integrating QKD into existing security protocols remains an unsolved challenge due to fundamental mismatches between pre-distributed quantum keys and computational key exchange paradigms. This paper presents the first systematic comparison of sequential versus parallel hybrid QKD-PQC key establishment strategies for IPsec, revealing fundamental protocol design principles that extend beyond specific implementations. We introduce two novel approaches for incorporating QKD into Internet Key Exchange version 2 (IKEv2) with support for both ETSI GS QKD 004 stateful and ETSI GS QKD 014 stateless API specifications: (1) a pure QKD approach that replaces computational key derivation with identifier-based quantum key coordination, and (2) a unified QKD-KEM abstraction that enables parallel composition of quantum and post-quantum cryptographic methods within existing protocol frameworks. Our key insight is that parallel hybrid approaches eliminate the multiplicative latency penalties inherent in sequential methods mandated by RFC 9370, achieving significant performance improvements under realistic network conditions. Performance evaluation using a Docker-based testing framework with IDQuantique QKD hardware demonstrates that the parallel hybrid approach significantly outperforms sequential methods under network latency conditions, while pure QKD achieves minimal bandwidth overhead through identifier-based key coordination. Our implementations provide practical quantum-enhanced IPsec solutions suitable for critical infrastructure deployments requiring defense-in-depth security.
\end{abstract}

\date{}

\section{Introduction}

The increasing anticipation of quantum computing capabilities renders traditional cryptographic methods vulnerable, demanding the exploration of quantum-resistant alternatives for secure communication protocols. In response to this quantum threat, the cryptographic community has developed two primary defense strategies: Post-Quantum Cryptography (PQC)~\cite{bernstein2009introduction}, which relies on mathematical problems believed to be hard even for quantum computers, and Quantum Key Distribution (QKD)~\cite{bennett2014quantum, gisin2002quantum}, which provides information-theoretic security based on fundamental quantum mechanical principles. The recent standardization of ML-KEM (Module-Lattice-Based Key Encapsulation Mechanism) as FIPS 203~\cite{fips203} by NIST in August 2024 establishes the primary post-quantum key establishment standard that organizations will deploy for quantum-resistant communications. While ML-KEM provides computational security against quantum attacks, hybrid approaches that combine standardized post-quantum algorithms with QKD could offer defense-in-depth strategies particularly valuable for critical infrastructure deployments.

IPsec (Internet Protocol Security) is a widely deployed secure communication protocol suite that provides security at the network layer, through two main protocols: Authentication Header (AH) and Encapsulating Security Payload (ESP), to handle data integrity, authentication, and encryption. These protocols require cryptographic keys to function, which are exchanged using the IKE (Internet Key Exchange) protocol. IPsec's architecture was first standardized in RFC 2401~\cite{rfc2401}, later updated in RFC 4301~\cite{rfc4301}, while key management evolved from IKEv1~\cite{rfc2409} and ISAKMP (Internet Security Association and Key Management Protocol)~\cite{rfc2408} to the initial IKEv2 specification~\cite{rfc4306}, and finally to its current form in RFC 7296~\cite{rfc7296}. To ensure consistent implementation and interoperability, standardized cryptographic profiles have been defined for IPsec, most notably the Suite B Cryptographic Suites specified in RFC 6379~\cite{rfc6379}, which formalized the use of elliptic curve cryptography and AES-GCM/GMAC for government and critical infrastructure applications. The Internet Key Exchange version 2 (IKEv2) is the protocol responsible for managing these keys, through security associations, and handling peer authentication.

Among the various IPsec implementations available, strongSwan~\cite{strongswan2024} has emerged as one of the most popular open-source solutions, offering support for both IKEv1 and IKEv2 protocols along with a flexible plugin architecture that facilitates the integration of novel cryptographic methods. It has been actively working on integration Post-Quantum Cryptography (PQC) support, as response to future quantum computing threats.
Nevertheless, integrating Quantum Key Distribution (QKD), which provides information-theoretic security based on quantum-physical principles, may impose operational requirements and current technical limitations.

Recent assessments from major cybersecurity agencies~\cite{QKD2024} highlight QKD's limited maturity, suggesting that hybrid approaches combining QKD with post-quantum cryptography could provide defense-in-depth for critical infrastructure scenarios where QKD is deemed necessary.

To address the practical challenge of transferring large cryptographic payloads such as post-quantum and QKD key material without causing problematic IP fragmentation, RFC 9242~\cite{rfc9242} introduces the \texttt{IKE\_INTERMEDIATE} exchange mechanism. This extension enables the secure transfer of large amounts of data between the initial \texttt{IKE\_SA\_INIT} and \texttt{IKE\_AUTH} exchanges, utilizing IKE-level fragmentation capabilities that are unavailable during the initial exchange. The \texttt{IKE\_INTERMEDIATE} exchange enables QKD integration by providing a mechanism to transfer large cryptographic payloads without IP fragmentation, which is particularly important when combining QKD with post-quantum algorithms that typically have large key sizes.

Building upon this foundation, RFC 9370~\cite{rfc9370} introduces a hybrid key exchange extension for IKEv2 that leverages the \texttt{IKE\_INTERMEDIATE} exchange mechanism to enable multiple sequential key exchanges within a single session. This approach allows the combination of classical (e.g., (EC)DH), post-quantum, and quantum-derived key material, where each exchange contributes to the overall key derivation process, ensuring security as long as at least one component remains unbroken. The protocol also introduces the \texttt{IKE\_FOLLOWUP\_KE} exchange for performing additional key exchanges during IKE SA rekeying or Child SA creation, providing support for hybrid cryptographic deployments.

Previously, RFC 8784~\cite{rfc8784} introduced a simpler method for mixing preshared keys into IKEv2 for post-quantum security, primarily protecting Child SAs but not the initial IKE SA. Building on this foundation, an active IETF draft~\cite{ietf-ipsecme-ikev2-qr-alt-08} proposes to extend RFC 8784 to enable the use of Post-quantum Preshared Keys (PPKs), which could be derived from QKD systems, in the \texttt{IKE\_INTERMEDIATE} and \texttt{CREATE\_CHILD\_SA} exchanges, addressing the challenge of leveraging fresh PPKs without reestablishing IKE SA from scratch while also protecting the initial IKE SA itself.

Additionally, in recent standardization efforts for other secure communication protocols, a framework for hybrid key exchange that combines traditional and post-quantum cryptography is being proposed~\cite{stebila2020hybrid}. This approach demonstrates how to achieve security through the concatenation of shared secrets from different key exchange methods, maintaining protection as long as at least one component remains unbroken. 

In this paper, we propose a hybrid quantum-resistant integration for IKEv2/IPsec, which is also suitable for other secure communication protocol such as TLS. Our approach shares conceptual similarities with the AQUA design by Nagayama and Van Meter~\cite{nagayama-ipsecme-ipsec-with-qkd-01}, which proposed exchanging QKD key identifiers instead of Diffie-Hellman values in IKEv2. Though this draft has since expired without proceeding to standardization, its core concept remains valuable. We extend this concept in two complementary ways: firstly, by integrating it with the RFC 9370 hybrid key exchange framework to enable sequential composition of QKD-derived keys with post-quantum cryptographic primitives; and second, by implementing a unified hybrid QKD-KEM Key Establishment (KE) approach that encapsulates QKD key management in a KEM-like interface. These strategies represent different design philosophies: the sequential approach maintains clear separation between cryptographic methods while the hybrid QKD-KEM offers tighter integration with reduced protocol overhead. Our strongSwan-based implementations support both ETSI GS QKD 004~\cite{etsi2020qkd} and ETSI GS QKD 014~\cite{etsi2020qkd014} API specifications, addressing interoperability challenges with commercial QKD hardware systems.

The remainder of this paper is organized as follows: Section~\ref{sec:related_work} provides an overview of the related work in the field for both IPsec and TLS. Section~\ref{sec:architecture} outlines the proposed system architecture, which includes different approaches. In section~\ref{sec:security}, a security analysis of the proposed approaches is carried out.
Section~\ref{sec:implementation} gives implementation details and Section~\ref{sec:experimental_setup} describes experimental setup for testing and evaluation. Section~\ref{sec:results} presents the results and Section~\ref{sec:discussion} discusses their significance from a comparative point of view. Finally, Section~\ref{sec:conclusions} concludes the paper and outlines directions for future research.

\section{Related Work}
\label{sec:related_work}

The integration of QKD with established security communication protocols has evolved significantly over the past two decades, with approaches targeting both IPsec and TLS frameworks. These developments have progressed from simple adaptations to sophisticated hybrid implementations.

\subsection{QKD Integration with IPsec}

Early research by Sfaxi et al. \cite{sfaxi2005using} introduced SeQKEIP, which added a preliminary quantum key exchange phase to the ISAKMP framework, deliberately bypassing IKE compatibility issues. Berzanskis et al. \cite{berzanskis2009method} took a different approach with a hybrid model that combined classical IKE and QKD-derived keys through XOR operations, requiring both systems to be compromised for a successful attack.

The AQUA approach by Nagayama and Van Meter \cite{nagayama-ipsecme-ipsec-with-qkd-01} proposed IKEv2 extensions for QKD integration, introducing QKD KeyID and Fallback payloads to exchange identifiers for pre-shared quantum keys rather than transmitting Diffie-Hellman values during key exchange.

DARPA's quantum network implementation \cite{elliott2007darpa}, analyzed by Dervisevic and Mehic \cite{dervisevic2021overview}, employed a 'rapid-reseeding' approach where QKD-derived bits supplemented the traditional IKEv1 key material in the KEYMAT hash function. This technique was further refined by MagiQ Technologies through direct XOR combination of QKD and IKE-generated keys. These early implementations highlighted key synchronization and rekeying rates as critical challenges for practical deployment.

The European SECOQC project (2004-2008) demonstrated practical QKD-IPsec integration using VPN tunnels with AES encryption refreshed every 20 seconds via quantum keys, establishing the hop-by-hop paradigm for quantum-secured IPsec communications \cite{mehic2020quantum}.

Significant theoretical advancements came from Dowling et al. \cite{dowling2020many}, whose Muckle protocol provides a framework for Hybrid Authenticated Key Exchange applicable to IPsec environments. Their approach combines classical, post-quantum, and QKD-derived key material to create a system resilient against vulnerabilities in any single component. By utilizing QKD-generated pre-shared keys for authentication, Muckle avoids the computational overhead of post-quantum signature schemes while maintaining security guarantees. Similarly, Huang et al. \cite{huang2020practical} proposed a "triple-security" method specifically designed for protocols like IPsec, theoretically integrating classical, post-quantum, and quantum key materials.

Marksteiner and Maurhart~\cite{marksteiner2015protocol} proposed a fundamentally different approach, bypassing IKE entirely in favor of a specialized key synchronization protocol for QKD-secured IPsec connections. Their solution focused on achieving rapid key refresh rates (up to 50 changes per second) with a fault-tolerant design to maintain synchronization even under suboptimal network conditions. While our approach leverages the standardized IKEv2 framework through strongSwan, their work offers valuable insights on the performance and security implications of different key refresh strategies.

Recent standardization efforts have focused on practical integration strategies for quantum-resistant key material into IKEv2. RFC 8784~\cite{rfc8784} established the foundation by introducing a method to mix Post-quantum Preshared Keys (PPKs) into the IKEv2 protocol, primarily protecting Child SAs against future quantum threats. Building on this work, Smyslov \cite{ietf-ipsecme-ikev2-qr-alt-06} proposes an extension that enables the use of these PPKs in both \texttt{IKE\_INTERMEDIATE} and \texttt{CREATE\_CHILD\_SA} exchanges. The European Telecommunications Standards Institute has also recognized the strategic importance of QKD integration with IPsec, documenting specific use cases and deployment scenarios for this integration~\cite{etsi2010qkdusecases}.

From a network architecture perspective, Lopez et al.~\cite{lopez2019applying} introduced the Software Defined QKD Node (SDQKDN) concept that aligns QKD deployment with the Software-Defined Networking (SDN) paradigm. Their proposal enables programmable QKD networks that can be dynamically configured to support various security applications, including IPsec VPNs, through standardized interfaces. This architectural approach is particularly valuable for telecommunications providers deploying QKD in next-generation networks, where automated service provisioning and network function virtualization are increasingly important.

China's large-scale deployments, including the Beijing-Shanghai backbone and metropolitan networks in Jinan and Wuhan, have validated production QKD-IPsec implementations with VoIP and video conferencing services delivered through IPsec VPN protocols using quantum-derived keys \cite{mehic2020quantum}.

Recent commercial security solutions have begun implementing QKD-IPsec integration features. FortiOS 7.4.2 has integrated IPsec key retrieval from QKD systems using the ETSI standardized API~\cite{fortinet2024qkd}, enabling FortiGate firewalls to connect to a Key Management Entity (KME) to obtain quantum-derived keys for IPsec tunnels, with options to either allow or require QKD for each VPN connection. Similarly, Juniper Networks' SRX Series Firewalls now support quantum-safe IPsec VPNs~\cite{juniper2025qkd}, implementing RFC 8784 for mixing Post-quantum Preshared Keys (PPK) in IKEv2. Juniper's implementation interfaces with external QKD devices using ETSI QKD API specifications, and offers both static key profiles for testing environments and dynamic quantum key manager profiles for production deployments. These commercial implementations from leading security vendors illustrate the industry's movement toward standardized QKD integration with established security protocols.

A recent field trial by Sibson et al.~\cite{sibson2024field} demonstrated the practical deployment of this Juniper quantum-secured IPsec solution, using chip-based QKD over 28.5 km of deployed fiber. Their implementation with Juniper SRX4600 routers validated the RFC 8784 integration approach with keys supplied via the ETSI GS QKD 014 API.

Major cloud providers are also exploring QKD-secured IPsec for their infrastructure. AWS Center for Quantum Networking has conducted a field trial in Singapore connecting two sites separated by 3 km (16 km fiber distance) using FortiGate firewalls to create an IPsec tunnel consuming quantum-generated keys, integrated with AWS Edge Compute hardware~\cite{aws2023quantum}. This demonstrates growing industry interest in quantum-secured communication across the technology sector.
More recently, Alia et al. \cite{alia2024100} demonstrated an industrial-grade QKD-IPsec implementation achieving 100 Gbps throughput over 46 km of deployed fiber using ID Quantique hardware. Their approach employed a sequential hybrid key exchange model that first established Security Associations with traditional Diffie-Hellman before incorporating QKD-derived keys in the key derivation function for Child SAs, showing that quantum-enhanced IPsec can maintain production-level performance while refreshing keys every 120 seconds.

\subsection{QKD Integration with TLS}

Parallel efforts in TLS integration offer valuable insights applicable to our IPsec work, particularly in adapting existing frameworks to QKD's pre-shared key model. Early contributions laid the groundwork for such adaptations: Aguado et al.~\cite{aguado2017hybrid} modified Diffie-Hellman exchanges to incorporate quantum key identifiers in SSH and TLS protocols, enabling seamless integration while preserving protocol structure. Building on this, Rijsman et al.~\cite{rijsman2019openssl} pioneered QKD integration with OpenSSL by repurposing the engine infrastructure, allowing quantum key retrieval without modifying core protocol code and demonstrating how existing security frameworks could accommodate QKD's fundamentally different key distribution model.

Subsequent advancements have expanded on this direction, often incorporating hybrid architectures that combine QKD with PQC for enhanced resilience and multi-layer security. Kozlovičs et al.~\cite{kozlovivcs2023quantum} introduced a "virtual KEM" coordinating key retrieval via identifiers and hashes, leveraging PQC for key distribution center links and QKD for end-to-end encryption. Rubio-Garc\'ia et al. progressed through a multi-level design allocating QKD to fiber and PQC to wireless segments in a three-layer hybrid~\cite{garcia2023enhancing}, followed by a triple-hybrid TLS 1.3 scheme concatenating ECDH, CRYSTALS-Kyber, and QKD keys in the key schedule (incurring 68\% QKD-driven overhead but validating feasibility~\cite{garcia2023quantum}) and later QKD integration into OpenSSL 3.2.0 for 2.5 kbit/s generation with bandwidth-efficient 36-byte identifiers~\cite{garcia2024integrating}. Complementing these, Rencis et al.~\cite{rencis2024hybrid} proposed a "QKD as a Service" framework applying segment-specific cryptography for composite security.

Further hybrid PQC-QKD developments include Zeng et al.'s series and parallel designs, optimizing performance/security trade-offs with formal vulnerability analysis in quantum-classical networks~\cite{zeng2024practical}. Our prior TLS integration~\cite{blanco2025qkd} similarly unified QKD and PQC via OpenSSL's provider infrastructure, supporting ETSI QKD-004 and -014 for hardware interoperability and application compatibility. 

Recent pure QKD-focused integrations include Pr\'evost et al.~\cite{prevost2025etsi}, who designed a novel TLS variant ("TLS-QKD") compliant with ETSI GS QKD 014 and implemented it through Rustls modifications, replacing public-key handshakes with local KME retrieval while ensuring backward compatibility; their protocol features a challenge-response mechanism for proving quantum key possession, preventing replays, and enabling mutual verification, formally verified via ProVerif, with practicality demonstrated in a QKD-encrypted video conference over 25 km.

Current standardization efforts through ITU-T \cite{itu2024qkdtls} are working to formalize these approaches, particularly addressing the challenges of deployments where QKD and protocol endpoints may be separated by untrusted networks.

\section{System Architecture and Design}
\label{sec:architecture}

The integration of QKD into existing cryptographic frameworks presents fundamental architectural challenges that stem from different operational paradigms~\cite{alleaume2014using}. Traditional key exchanges and KEMs follow a computational model where shared secrets are actively derived from public values exchanged between parties. In contrast, QKD operates by pre-establishing keys through quantum channels, which are then referenced through identifiers during classical communication. This fundamental difference creates a mismatch in existing cryptographic APIs designed around the computational model.

Similar to how Rijsman et al. \cite{rijsman2019openssl} repurposed the Diffie-Hellman engine infrastructure, implementing pure QKD in modern cryptographic frameworks remains challenging due to architectural constraints. 
While the ETSI QKD API standards \cite{etsi2020qkd,etsi2020qkd014} provide well-defined interfaces for QKD operations, with ETSI GS QKD 004 defining the basic application interface and ETSI GS QKD 014 specifying the REST-based key delivery API, OpenSSL 3.x's \textbf{provider} API offers KEM primitives that, while suitable for classical and post-quantum key exchanges, do not naturally accommodate QKD's key distribution model. However, as we will show, strongSwan's plugin architecture provides a more suitable abstraction for pure QKD integration, as it allows direct modification of key exchange methods without having to adapt to existing key exchange primitives. A hybrid approach, combining QKD with classical or post-quantum KEMs, could also be feasible in OpenSSL as it maintains the existing key exchange abstractions.

Our architecture supports QKD-derived keys via standardized ETSI QKD-004 and QKD-014 APIs and aligns well with the Software-Defined Networking (SDN) paradigm. As Lopez et al.~\cite{lopez2019applying} describe, the centralized management of QKD resources through standardized interfaces enables the creation of programmable quantum-secure networks. By using these standardized APIs, it could be extended to interface with SDN controllers, allowing dynamic reconfiguration of IPsec security parameters based on network conditions and security requirements. This SDN-compatible approach is particularly valuable for telecommunications providers deploying QKD in next-generation networks, where automated service provisioning and network function virtualization are becoming increasingly important.

Our QKD integration with IKEv2 transforms the traditional key exchange mechanism by incorporating quantum-derived cryptographic material through standardized ETSI QKD API specifications. We propose two distinct integration strategies: pure QKD integration that replaces classical Diffie-Hellman exchanges entirely, and hybrid QKD-PQC integration that combines quantum keys with post-quantum cryptography for enhanced security.

\subsection{Pure QKD Integration}

The pure QKD integration modifies the fundamental IKEv2 key exchange by replacing computational key derivation with pre-distributed quantum keys. Unlike traditional IKEv2 where both peers contribute entropy through Diffie-Hellman exchange, our QKD adaptation establishes shared secrets through prior quantum key distribution, using the classical channel only to coordinate which pre-shared quantum key to use.

We define two operational flows depending on the ETSI QKD API specification and deployment requirements:

\paragraph{Client-Initiated Flow}

The client-initiated approach follows the following sequence:

\begin{enumerate}
   \item \textbf{Key Generation (Alice):} The IKE initiator's \texttt{get\_public\_key()} method retrieves a quantum key and corresponding identifier from the Key Management Entity (KME):
   \begin{itemize}
       \item ETSI 004: Establishes a session via \texttt{OPEN\_CONNECT()} and receives a key ID with starting index
       \item ETSI 014: Obtains key and ID using \texttt{GET\_KEY()}
   \end{itemize}
   
   \item \textbf{\texttt{IKE\_SA\_INIT} Request:} Alice transmits the QKD key identifier returned by \texttt{get\_public\_key()} in the KEY\_EXCHANGE payload instead of a Diffie-Hellman public value
   
   \item \textbf{Key Retrieval (Bob):} The IKE responder's \texttt{set\_public\_key()} method receives the key identifier and retrieves the corresponding quantum key:
   \begin{itemize}
       \item ETSI 004: Establishes session via \texttt{OPEN\_CONNECT(key\_id)} then calls \texttt{GET\_KEY()}
       \item ETSI 014: Uses \texttt{GET\_KEY\_WITH\_IDS(key\_id)}
   \end{itemize}
   
   \item \textbf{\texttt{IKE\_SA\_INIT} Response:} Bob's \texttt{get\_public\_key()} returns an empty payload as no additional key material needs exchange
   
    \item \textbf{Key Derivation (Alice):} Alice's \texttt{set\_public\_key()} processes the empty response. For ETSI 004, this triggers a \texttt{GET\_KEY()} call to retrieve the quantum key material using the previously established session and key ID. For ETSI 014, Alice already has the key from step 1.
   
   \item \textbf{Shared Secret Computation:} Both parties' \texttt{get\_shared\_secret()} methods return the retrieved quantum key material for IKE SA key derivation
\end{enumerate}

\begin{figure*}[htbp]
    \centering
    \includegraphics[width=1\textwidth]{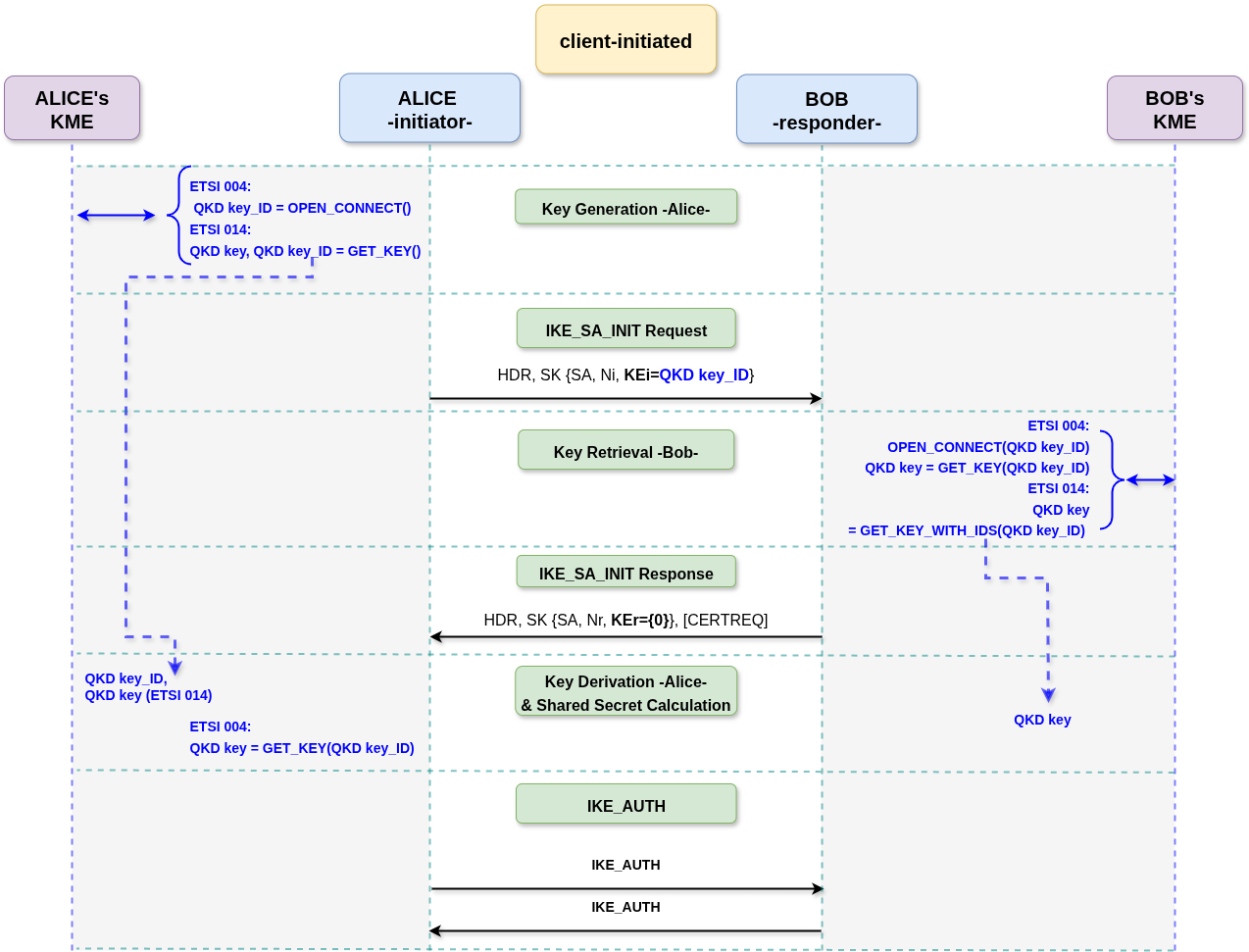}
    \caption{IKEv2 handshake with QKD integration using the client-initiated approach. The diagram shows the protocol flow for both ETSI 004 and ETSI 014 specifications, illustrating how quantum key identifiers replace Diffie-Hellman values in the \texttt{IKE\_SA\_INIT} exchange and the parallel communication with Key Management Entities for quantum key retrieval.}
    \label{fig:ike_v2_qkd}
\end{figure*}

Figure~\ref{fig:ike_v2_qkd} illustrates the complete protocol flow for the client-initiated approach, showing the parallel communication between IPsec endpoints and their respective KMEs.

\paragraph{Server-Initiated Flow}

The server-initiated approach reverses the key generation responsibility:

\begin{enumerate}
   \item \textbf{\texttt{IKE\_SA\_INIT} Request:} Alice signals QKD support without retrieving keys, sending an empty KEY\_EXCHANGE payload
   
   \item \textbf{Key Generation (Bob):} The IKE responder establishes a session and retrieves a quantum key:
   \begin{itemize}
       \item ETSI 004: Calls \texttt{OPEN\_CONNECT()} with NULL Key\_stream\_ID to obtain a new Key\_stream\_ID, then immediately calls \texttt{GET\_KEY()} to retrieve the quantum key material
       \item ETSI 014: Calls \texttt{GET\_KEY()} to obtain key and ID
   \end{itemize}
   
   \item \textbf{\texttt{IKE\_SA\_INIT} Response:} Bob transmits the QKD key identifier in the KEY\_EXCHANGE payload
   
   \item \textbf{Key Retrieval (Alice):} The IKE initiator establishes the session and retrieves the corresponding quantum key:
   \begin{itemize}
       \item ETSI 004: Calls \texttt{OPEN\_CONNECT(key\_id)} to join the stream, then \texttt{GET\_KEY()} to obtain the key material
       \item ETSI 014: Calls \texttt{GET\_KEY\_WITH\_IDS(key\_id)}
   \end{itemize}
   
   \item \textbf{Key Derivation:} Both parties establish shared IKE SA keys from the quantum material
\end{enumerate}

Unlike traditional IKEv2, where both peers contribute entropy to the shared secret through Diffie-Hellman exchange, our QKD adaptation establishes the shared secret through prior quantum key distribution, with the classical channel used only to coordinate which pre-shared quantum key to use.

For ETSI 004 implementations, the transmitted key identifier payload may include additional bytes beyond the 16-byte Key\_stream\_ID (KSID) to specify the starting index (a 32-bit unsigned integer, typically 4 bytes). This allows precise positioning within the key stream for synchronization, particularly useful when retrieving multiple key chunks from the same established session.

\subsubsection{Key Stream Management Considerations}

Although our current implementation simplifies key retrieval by always using the first index and generating a new KSID for each session, the stateful nature of ETSI 004 enables more advanced session maintenance across multiple key retrievals using a single KSID, with the index parameter shifted (incremented) for each \texttt{GET\_KEY()} call to access subsequent chunks. This advanced approach can be advantageous for IPsec, where a long-lived IKE SA can maintain one QKD session and use different indices for Child SA rekeying, minimizing protocol overhead. Keeping the KSID constant while incrementing the index avoids repeated \texttt{OPEN\_CONNECT()} calls; deployments using such advanced management must handle session lifetime (via QoS TTL) and \texttt{CLOSE()} calls.

\subsection{Hybrid QKD-PQC Integration}

Beyond pure QKD integration, we implement hybrid KEs that combine QKD with PQC to provide defense-in-depth security. These hybrid methods address concerns raised by cybersecurity agencies regarding QKD's operational maturity while leveraging its information-theoretic security guarantees.

\begin{table*}[htbp]
\centering
\caption{Cryptographic payload sizes for QKD+ML-KEM hybrid KEs}
\label{tab:qkd_pqc_payload_sizes}
\resizebox{\textwidth}{!}{%
\begin{tabular}{|l|c|c|c|c|}
\hline
\multirow{3}{*}{\textbf{Hybrid}} & \multicolumn{2}{c|}{\textbf{Client-Initiated}} & \multicolumn{2}{c|}{\textbf{Server-Initiated}} \\
\multirow{3}{*}{\textbf{KE}} & \multicolumn{2}{c|}{\textbf{Approach}} & \multicolumn{2}{c|}{\textbf{Approach}} \\
\cline{2-5}
& \textbf{Request} & \textbf{Response} & \textbf{Request} & \textbf{Response} \\
& \textbf{(bytes)} & \textbf{(bytes)} & \textbf{(bytes)} & \textbf{(bytes)} \\
\hline
QKD + & 816 & 768 & 800 & 784 \\
ML-KEM-512 & (16 ID [+4 IDX opt. ETSI 004] + 800 PK) & (768 CT) & (800 PK) & (16 ID [+4 IDX opt. ETSI 004] + 768 CT) \\
\hline
QKD + & 1200 & 1088 & 1184 & 1104 \\
ML-KEM-768 & (16 ID [+4 IDX opt. ETSI 004] + 1184 PK) & (1088 CT) & (1184 PK) & (16 ID [+4 IDX opt. ETSI 004] + 1088 CT) \\
\hline
QKD + & 1584 & 1568 & 1568 & 1584 \\
ML-KEM-1024 & (16 ID [+4 IDX opt. ETSI 004] + 1568 PK) & (1568 CT) & (1568 PK) & (16 ID [+4 IDX opt. ETSI 004] + 1568 CT) \\
\hline
\end{tabular}%
}
\vspace{0.5em}
\captionsetup{font=footnotesize}
\caption*{Cryptographic payload sizes exclude IKE protocol headers and overhead. ID = QKD Key Identifier (16 bytes), IDX = optional starting index (4 bytes, 32-bit unsigned integer) for ETSI 004, PK = ML-KEM Public Key, CT = ML-KEM Ciphertext. Client-initiated: initiator sends QKD Key ID + ML-KEM PK; responder sends ML-KEM CT only. Server-initiated: initiator sends ML-KEM PK only; responder sends QKD Key ID + ML-KEM CT. Totals shown without optional IDX; add 4 bytes to the payload containing the ID when using explicit index in ETSI 004 for precise stream positioning.}
\end{table*}

Our hybrid approach encapsulates both QKD and post-quantum operations within a unified KEM interface, maintaining the original IKEv2 message exchange pattern. Similar to pure QKD integration, we support both client-initiated and server-initiated flows, with the same ETSI API compatibility constraints:

\paragraph{Client-Initiated Hybrid Flow}
The client-initiated hybrid approach, compatible with both ETSI specifications, follows this sequence:

\begin{enumerate}
    \item \textbf{Hybrid Key Generation (Alice):} The initiator simultaneously generates a post-quantum keypair and retrieves a QKD key identifier from the KME via the appropriate ETSI API
    \item \textbf{\texttt{IKE\_SA\_INIT} Request:} Alice transmits both the post-quantum public key and QKD key identifier in a single KEY\_EXCHANGE payload
    \item \textbf{Hybrid Encapsulation (Bob):} The responder retrieves the corresponding quantum key using the received identifier and performs post-quantum encapsulation with the public key
    \item \textbf{\texttt{IKE\_SA\_INIT} Response:} Bob returns the post-quantum ciphertext in the KEY\_EXCHANGE payload
    \item \textbf{Hybrid Decapsulation (Alice):} The initiator performs post-quantum decapsulation and retrieves the quantum key using the previously obtained identifier
    \item \textbf{Key Derivation:} Both parties concatenate the shared secrets: \texttt{shared\_secret = PQC\_secret | QKD\_key}
\end{enumerate}

\paragraph{Server-Initiated Hybrid Flow}

The server-initiated hybrid approach reverses the QKD key generation responsibility while maintaining the post-quantum key exchange pattern:

\begin{enumerate}
    \item \textbf{PQC Key Generation (Alice):} The initiator generates only a post-quantum keypair
    \item \textbf{\texttt{IKE\_SA\_INIT} Request:} Alice transmits the post-quantum public key in the KEY\_EXCHANGE payload
    \item \textbf{Hybrid Encapsulation (Bob):} The responder establishes a QKD session to retrieve a key identifier:
    \begin{itemize}
        \item ETSI 004: Calls \texttt{OPEN\_CONNECT()} with NULL to obtain Key\_stream\_ID, then calls \texttt{GET\_KEY()} to retrieve the quantum key material
        \item ETSI 014: Calls \texttt{GET\_KEY()} to obtain key and ID
    \end{itemize}
    Bob then performs post-quantum encapsulation and combines both the ciphertext and QKD identifier
    \item \textbf{\texttt{IKE\_SA\_INIT} Response:} Bob returns both the post-quantum ciphertext and QKD key identifier in the \texttt{KEY\_EXCHANGE} payload
    \item \textbf{Hybrid Decapsulation (Alice):} The initiator performs post-quantum decapsulation and retrieves the corresponding quantum key:
    \begin{itemize}
        \item ETSI 004: Calls \texttt{OPEN\_CONNECT(key\_id)} to join the stream, then \texttt{GET\_KEY()} to obtain the key material
        \item ETSI 014: Calls \texttt{GET\_KEY\_WITH\_IDS(key\_id)}
    \end{itemize}
    \item \textbf{Key Derivation:} Both parties establish the concatenated shared secret from both cryptographic methods
\end{enumerate}

Table~\ref{tab:qkd_pqc_payload_sizes} illustrates the cryptographic payload sizes for QKD+ML-KEM hybrid key exchanges as concrete examples of our implementation. The client-initiated approach places the QKD key identifier alongside the post-quantum public key in the request, while the server-initiated approach includes the identifier with the ciphertext in the response.

Both hybrid flows are supported through our QKD-KEM plugin, which integrates both ETSI 004 and ETSI 014 specifications via our unified C wrapper. The QKD-KEM provider automatically handles the appropriate ETSI API calls based on the configured backend, enabling seamless operation across different QKD hardware platforms while maintaining the hybrid security model.

\section{Security Analysis}
\label{sec:security}

This section analyzes the security properties of our QKD integration approaches, examining both pure QKD and hybrid schemes under different threat models.

\subsection{QKD Integration Security (Pure and Hybrid)}

Our QKD integration approaches share common security properties and vulnerabilities regardless of whether they operate in pure QKD or hybrid QKD-PQC modes, as both rely on the same underlying ETSI API mechanisms and protocol flows.

\textbf{QKD Key Security:} The quantum-generated keys provide information-theoretic security against computationally unlimited adversaries, including quantum computers. However, this security relies on correct QKD implementation and the absence of side-channel attacks on the physical hardware~\cite{lydersen2010hacking, QKD2024}. Our approach inherits these properties directly from the underlying QKD system.

\textbf{Key Identifier Transmission:} QKD key identifiers transmitted in \texttt{IKE\_SA\_INIT} messages do not compromise security as they reference pre-shared keys rather than enabling key derivation. An adversary intercepting identifiers cannot reconstruct the corresponding quantum keys without access to the KME. However, identifier exposure could enable traffic analysis or denial-of-service attacks by requesting the same keys.

\textbf{Protocol Vulnerabilities:} The main security risk occurs during the transition period between \texttt{IKE\_SA\_INIT} and \texttt{IKE\_AUTH} exchanges. Until peer authentication completes, an active adversary could potentially mount person-in-the-middle attacks by substituting key identifiers, causing endpoints to derive different shared secrets. Such attacks would manifest as authentication failures during the IKE\_AUTH exchange, as peers cannot verify each other's AUTH payloads when using mismatched shared secrets, though distinguishing malicious substitution from legitimate protocol failures may be challenging. Additionally, while QKD provides information-theoretic security for key establishment, peer authentication still relies on asymmetric cryptography (RSA, ECDSA signatures, X.509 certificates, or post-quantum alternatives like ML-DSA/Dilithium), creating a fundamental limitation where overall security remains bounded by the quantum vulnerability of classical authentication mechanisms unless post-quantum signature schemes are employed~\cite{alleaume2014using, QKD2024}.

\textbf{Forward Secrecy:} QKD integration can provide forward secrecy properties comparable to traditional IKEv2, where perfect forward secrecy requires that endpoints forget not only the session keys but also any information that could be used to recompute those keys~\cite{rfc7296}. Each session uses distinct quantum keys, and compromise of long-term authentication credentials does not reveal previously established session keys, provided the quantum key material has been properly erased from QKD storage systems~\cite{mehic2020quantum}.

\textbf{Denial-of-Service Vulnerabilities:} QKD integration introduces DDoS attack vectors that exploit the finite nature of quantum key resources~\cite{dervisevic2022simulations, mehic2020quantum}. Attackers can exhaust quantum key pools by initiating multiple handshakes without completing authentication, depleting resources faster than they can be replenished~\cite{mehic2022tackling}. Additionally, QKD networks are vulnerable to physical-layer attacks that can disrupt key generation and require real-time monitoring and mitigation strategies~\cite{hugues2019monitoring}. The blocking nature of ETSI API calls enables amplification attacks where malicious requests force legitimate endpoints to wait for expensive KME operations. These vulnerabilities require mitigation through rate limiting, key reservation mechanisms, and proper timeout handling in production deployments.

\subsection{Hybrid QKD-PQC Specific Security}

Our hybrid implementations concatenate QKD-derived keys with post-quantum shared secrets: $\text{shared\_secret} = \text{PQC\_secret} \| \text{QKD\_key}$. This parallel composition provides security as long as at least one component remains unbroken.

\textbf{Composite Security Model:} The hybrid approach achieves defense-in-depth by combining computational security (post-quantum algorithms) with information-theoretic security (QKD). An adversary must simultaneously break both the post-quantum algorithm and compromise the QKD system to recover the shared secret. This addresses concerns about QKD implementation vulnerabilities while providing quantum-resistant fallback protection.

\textbf{Key Concatenation Security:} The security of concatenated keys follows established hybrid cryptography principles as described in the TLS hybrid design framework~\cite{stebila2020hybrid}. Under the assumption that shared secrets are fixed length once the combination is fixed, our concatenation approach corresponds to established combiners that preserve security as long as at least one component key exchange remains secure. The resulting shared secret maintains the security properties of the strongest component algorithm.

\textbf{Attack Surface Analysis:} Hybrid implementations present a larger attack surface due to multiple cryptographic components, but failures in one system do not compromise overall security. QKD vulnerabilities (side-channel attacks, implementation flaws) are mitigated by post-quantum security, while potential future quantum attacks on post-quantum algorithms are addressed by information-theoretic QKD security.

\section{Implementation}
\label{sec:implementation}

Our implementation extends strongSwan~\cite{strongswan2024} with two custom plugins that integrate quantum-resistant cryptography into IPsec. While strongSwan 6.0.0 introduced native support for RFC 9370's multiple key exchange mechanism and ML-KEM algorithms~\cite{strongswan2024_6_0_0}, our work focuses specifically on QKD integration through standardized ETSI APIs. Both plugins implement strongSwan's \texttt{key\_exchange\_t} interface, enabling seamless integration with existing IKEv2 protocol handlers and compatibility with strongSwan's RFC 9370 framework for sequential hybrid combinations.

\begin{figure}[htbp]
    \centering
    \includegraphics[width=1\columnwidth]{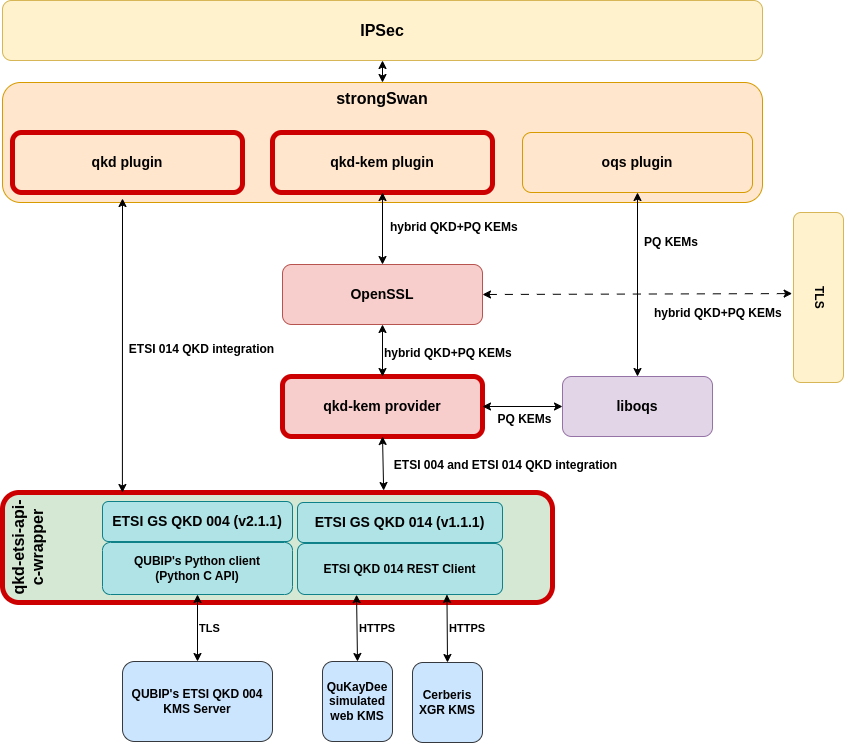}
    \caption{Software architecture of our QKD-IPsec integration. The diagram shows the interaction between strongSwan plugins, the QKD ETSI API C wrapper, and different QKD backends including QUBIP's Python-based KMS for ETSI 004 and native REST clients for ETSI 014.}
    \label{fig:qkd_IPsec_architecture}
\end{figure}

Figure~\ref{fig:qkd_IPsec_architecture} illustrates the complete software architecture, showing the interaction between our plugins, the ETSI API wrapper, and different QKD backends. The implementation is available in our public repository~\cite{qkd-plugins-strongswan}.

\begin{enumerate}
   \item \textbf{QKD Plugin}: Enables direct integration of quantum key material into IKEv2
   \item \textbf{QKD-KEM Plugin}: Provides hybrid quantum/post-quantum security through a unified KEM interface
\end{enumerate}

\subsection{QKD Plugin}

The QKD plugin enables direct integration of quantum key material into IKEv2 through pure QKD key exchange. The plugin consists of two core components:

\begin{itemize}
   \item \textbf{Key Exchange Handler} (\texttt{qkd\_kex.c}): Implements the IKEv2 key exchange interface with QKD-specific operations. The \texttt{get\_public\_key()} method returns quantum key identifiers for initiators or empty values for responders. The \texttt{set\_public\_key()} method processes received identifiers and triggers quantum key retrieval from the KME. The \texttt{get\_shared\_secret()} method returns the retrieved quantum key material for IKE SA derivation.
   
   \item \textbf{ETSI API Adapter} (\texttt{qkd\_etsi\_adapter.c}): Provides an abstraction layer for communicating with QKD hardware through our ETSI API wrapper~\cite{blanco2024qkd}, supporting both ETSI GS QKD 004 and ETSI GS QKD 014 specifications.
\end{itemize}

\subsection{QKD-KEM Plugin}

The QKD-KEM plugin implements hybrid quantum/post-quantum security through a unified KEM interface. This approach maintains the original IKEv2 message exchange pattern while combining post-quantum algorithms with quantum key material. The plugin leverages our custom OpenSSL provider~\cite{qursa2024qkd} that hybridizes both ETSI standards, adapting concepts from our previous TLS integration work~\cite{blanco2025qkd}.

The \texttt{qkd\_kem.c} handler implements OpenSSL-based KEM operations where \texttt{get\_public\_key()} generates post-quantum keypairs or returns hybrid ciphertexts, \texttt{set\_public\_key()} performs encapsulation or decapsulation with integrated QKD key retrieval, and \texttt{get\_shared\_secret()} returns the concatenated shared secret combining both cryptographic methods.

\begin{itemize}
   \item \textbf{KEM Operations Handler} (\texttt{qkd\_kem.c}): Implements the OpenSSL-based key exchange operations:
   \begin{itemize}
       \item \texttt{get\_public\_key()}: For initiators, generates a keypair and returns the public key; for responders, returns the ciphertext containing the encapsulated shared secret
       \item \texttt{set\_public\_key()}: For initiators, processes the received ciphertext through decapsulation; for responders, encapsulates a shared secret using the received public key
       \item \texttt{get\_shared\_secret()}: Returns the established shared secret for IKE SA derivation
   \end{itemize}
\end{itemize}

The ETSI interface enables integration with QKD hardware through standardized API calls.

\subsection{QKD ETSI API C Wrapper}

To ensure compatibility across different QKD systems, we developed a unified C wrapper library~\cite{blanco2024qkd} implementing both ETSI QKD API specifications. The wrapper abstracts the fundamental differences between ETSI GS QKD 004's stateful stream-based interface (\texttt{OPEN\_CONNECT}, \texttt{GET\_KEY}, \texttt{CLOSE}) and ETSI GS QKD 014's stateless REST-based operations (\texttt{GET\_KEY}, \texttt{GET\_KEY\_WITH\_IDS}).

For the ETSI QKD-004 standard, we integrated QUBIP's implementation~\cite{qubip2024etsi}, which consists of a Python-based Key Management System (KMS) server and a client SDK implementing the protocol specification. Rather than rewriting this functionality in C, we developed a hybrid approach: our C wrapper library embeds the QUBIP Python client directly via the Python C API. The wrapper initializes a Python interpreter at runtime, imports the client module, and establishes a bidirectional bridge between data structures. When strongSwan calls our standard C functions, the wrapper translates these into Python method invocations, handling conversion of QoS parameters, key identifiers, and cryptographic material between languages. The Python client then manages the TLS-secured connection to the KMS server, implementing the ETSI protocol operations (\texttt{OPEN\_CONNECT} , \texttt{GET\_KEY}, \texttt{CLOSE}). This architecture preserves all the protocol-specific functionality of QUBIP's implementation while presenting a consistent, native C interface to our IPsec framework.

For the ETSI QKD-014 standard, our wrapper provides a native C implementation of the REST-based key delivery API, enabling connection with both hardware QKD systems and cloud-based simulators. The implementation features a modular backend approach that supports multiple platforms: IDQuantique Cerberis XGR and QuKayDee cloud simulator, with configuration-time selection through conditional compilation. The code uses libcurl for HTTPS requests with mutual TLS authentication and the jansson library for JSON processing. Each backend implements three primary functions: \texttt{get\_status} to query system capabilities, \texttt{get\_key} to request new quantum keys from the KME, and \texttt{get\_key\_with\_ids} to retrieve keys using previously shared identifiers. The implementation automatically handles certificate management for different user roles (initiator or responder) through environment variables, enabling proper authentication with QKD Key Management Entities. This architecture allows us to test with simulated environments during development and transition seamlessly to hardware QKD deployments without application code changes.

The unified wrapper allows identical IPsec configurations to work with different QKD backends without code changes.

\subsection{Implementation Security Considerations}

The security of our QKD integration depends significantly on implementation details across multiple software components. This work provides a proof-of-concept demonstration rather than a production-ready secure implementation, and the implementation security considerations are similar to those of other cryptographic software such as strongSwan plugins, OpenSSL providers, and ETSI API implementations.

\textbf{Memory Management:} Secure handling of quantum key material requires proper memory allocation, zeroing of sensitive data after use, and protection against memory disclosure attacks. Our strongSwan plugins follow standard practices for cryptographic key handling, but production deployments require additional hardening measures such as memory locking and secure deletion protocols.

\textbf{ETSI API Security:} The security boundary extends to the ETSI API implementations and the communication channels between IPsec endpoints and KMEs. Our C wrapper abstracts these interfaces but inherits the security properties of the underlying implementations, including QUBIP's Python-based client for ETSI 004 and native REST clients for ETSI 014. Production deployments must ensure these communication channels are secure.

\textbf{Timing and Side-Channel Resistance:} QKD key retrieval operations involve network communications and cryptographic computations that may introduce timing variations. While our implementation does not specifically address timing attack mitigation, production systems should implement constant-time operations and consider the timing implications of ETSI API calls.

\textbf{Network-Level Security Dependencies:} Our approach assumes secure channels between IPsec endpoints and their respective KMEs, as specified in ETSI standards. Deployments where KMEs are separated from IPsec endpoints by untrusted networks require additional protection mechanisms beyond our protocol modifications, such as dedicated secure links or additional encryption layers for KME communications.

\section{Experimental Setup}
\label{sec:experimental_setup}

To validate our QKD-IPsec integration, we developed a testing environment combining both physical QKD infrastructure and containerized software components~\cite{qkd-ipsec-docker}. Our experimental approach encompasses hardware QKD nodes, Docker-based network simulation, and automated measurement tools to provide controlled and reproducible testing conditions.

\subsection{Physical QKD Infrastructure}

\begin{figure}[htbp]
    \centering
    \includegraphics[width=1\columnwidth]{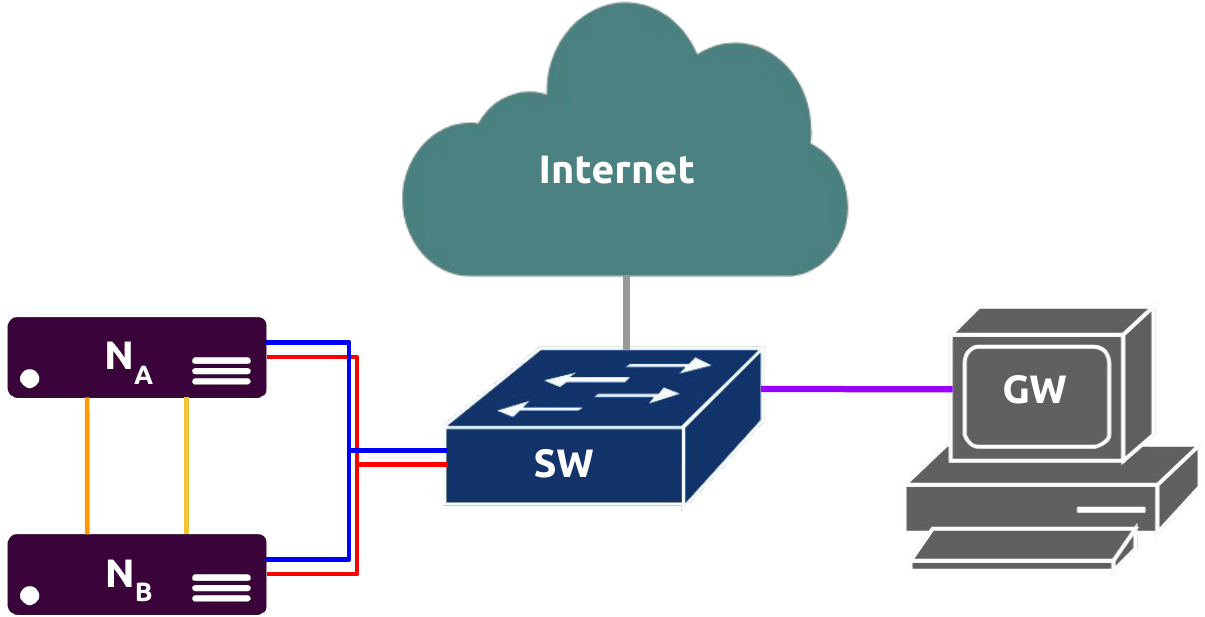}
    \caption{The setup consists of two IDQuantique Cerberis XGR nodes (N\_A and N\_B) connected via dedicated dark fiber for the quantum channel and a separate classical channel. Both nodes connect to a network switch (SW) that provides access to the Internet cloud and a gateway (GW) machine. The GW serves as the management endpoint where Docker containers (Alice and Bob) are deployed for the VPN testing environment.}
    \label{fig:qkd_network_topology}
\end{figure}

Our physical testbed is built around two IDQuantique Cerberis XGR QKD nodes (Figure~\ref{fig:qkd_network_topology}), labeled N\_A and N\_B, which establish a quantum channel over dedicated dark fiber while maintaining a separate classical communication channel. These QKD nodes connect to a central network switch (SW) that bridges the quantum network with both Internet connectivity and a dedicated gateway (GW) server. The GW hosts our Docker-based testing environment where the Alice and Bob containers are deployed, allowing them to request quantum keys from their respective QKD nodes while maintaining traditional network connectivity. This architecture physically separates the quantum key generation infrastructure from the IPsec tunnel endpoints, reflecting a realistic deployment scenario where QKD hardware serves as specialized key management entities that interface with conventional networking equipment.

\subsection{Containerized Testing Environment}

Our testbed architecture follows a client-server model with Alice functioning as the VPN client and Bob as the VPN server, both implemented as Docker containers to ensure experimental reproducibility and isolation. This containerized approach, based on a modified version of the strongX509 containerized testing environment \cite{strongx509-docker}, enables automated testing across multiple cryptographic proposals while maintaining consistent baseline conditions.

The Docker environment consists of a base strongSwan image containing our QKD and QKD-KEM plugins, with two derived containers configured for distinct roles and network topologies. The base image is built with configurable support for both plugin types, enabling runtime selection of different ETSI API backends (simulated or hardware) and QKD initiation modes (client or server-initiated flows). Build-time parameters control plugin compilation, QKD backend selection, and strongSwan version, allowing reconfiguration for different test scenarios.

Alice and Bob containers are deployed on separate virtual networks simulating a typical VPN gateway scenario. Alice operates on the internet network (172.30.0.0/24) representing remote clients, while Bob bridges both internet and intranet networks (172.31.0.0/16) acting as a VPN gateway providing secure access to internal resources.

The testing environment provides shared configuration volumes for cryptographic proposal definitions, certificates for QKD node authentication, and dedicated output directories for capturing experimental results. A centralized proposal configuration system enables automated testing across multiple cryptographic combinations without manual reconfiguration between test iterations. Each container maintains separate strongSwan configurations while sharing common cryptographic parameters through mounted volumes, ensuring consistent security associations across test runs.

The system is configured to load the QKD or QKD-KEM plugins alongside traditional cryptographic modules, allowing for the specification of quantum-enhanced cryptographic proposals in IKEv2 negotiations. While the peers authenticate using conventional X.509 certificates, the key exchange process is modified to utilize quantum-derived keys.

The automated testing framework implements a synchronization protocol between Alice and Bob containers, coordinating handshake initiation, data collection, and cleanup operations. Alice serves as the test controller, managing iteration counts and proposal selection, while Bob responds to connection requests and maintains consistent responder behavior. This orchestration ensures reproducible measurements across different network conditions and cryptographic proposals.

\subsection{Network Emulation}

To evaluate performance under realistic network conditions, we incorporated network emulation using the Pumba chaos testing tool \cite{alexei-led2024pumba}. This approach allows controlled introduction of network impairments between Alice and Bob containers while preserving their communication with QKD key management entities. Traffic control rules using Pumba's netem commands target only IPsec tunnel traffic, enabling isolated testing of protocol performance under varying latency, jitter, and packet loss conditions without affecting QKD backend operations.

\subsection{strongSwan Configuration}

Our strongSwan configuration was optimized for post-quantum cryptographic algorithms and QKD integration, addressing the inherently larger packet sizes required by these protocols. Initial experiments with default strongSwan parameters encountered packet processing failures due to insufficient buffer limits, manifesting as ``receive buffer too small, packet discarded'' errors when processing algorithms such as HQC-256 and FrodoKEM variants. These failures occurred because post-quantum algorithms generate significantly larger cryptographic payloads compared to classical algorithms (HQC-256 produces public keys of 7,245 bytes and ciphertexts of 11,424 bytes, while FrodoKEM-1344 generates public keys up to 21,520 bytes). When encapsulated within \texttt{IKE\_SA\_INIT} messages, packets exceeded 10,000 bytes as observed in our experimental logs.

To accommodate these requirements while maintaining realistic network parameters, we modified several strongSwan 6.0 default values~\cite{strongswan2024docs}. The default \texttt{max\_packet} limit of 10,000 bytes was increased to 25,000 bytes to accommodate complete post-quantum IKE messages before fragmentation, providing sufficient headroom for the largest expected cryptographic payloads while remaining within reasonable jumbo frame capabilities. For fragmentation control in our IPv4-based Docker testbed (172.30.0.0/24 and 172.31.0.0/16 networks), we configured \texttt{fragment\_size = 1472} bytes, overriding the conservative default of 1,280 bytes. This parameter controls the maximum size of individual IKE fragments when strongSwan divides large messages for transmission, representing the optimal fragment size for standard Ethernet networks (1500-byte MTU minus 20-byte IP header and 8-byte UDP header). Although not the most conservative choice, this setting reflects realistic network conditions and increases bandwidth efficiency for post-quantum key exchanges without compromising broad compatibility with standard networking infrastructure. The configuration also includes reduced timeout values (\texttt{half\_open\_timeout = 20} seconds vs. default 30, \texttt{retransmit\_timeout = 3} seconds vs. default 4, \texttt{retransmit\_tries = 3} vs. default 5) to accelerate experimental iterations while maintaining sufficient time for quantum key retrieval operations via ETSI API calls.

\subsection{Measurement Methodology}

All measurements are performed from Alice's perspective as the IKE initiator, providing consistent timing baselines across different cryptographic proposals. The measurement framework captures both network-level metrics through packet analysis and plugin-level performance through instrumented timing logs.

To evaluate the protocol overhead we conducted packet capture and analysis during the cryptographic handshake process. Network traffic was captured using \texttt{tcpdump} during the execution of each cryptographic proposal, generating PCAP files containing the complete communication exchange between Alice (initiator) and Bob (responder).

The captured traffic was subsequently processed using \texttt{tshark} with ISAKMP filtering to isolate IKEv2 protocol messages, excluding all non-IKE traffic such as TCP control messages, DNS queries, and ICMPv6 router advertisements. For each PCAP file, we extracted the following packet-level information: frame timestamp, IKE exchange type, message identifier, and frame length in bytes. The IKE exchange types analyzed include \texttt{IKE\_SA\_INIT} (type 34), \texttt{IKE\_AUTH} (type 35), \texttt{IKE\_INTERMEDIATE} (type 43), \texttt{INFORMATIONAL} (type 37), and \texttt{CREATE\_CHILD\_SA} (type 36).

Byte consumption analysis was performed at two levels: IKE-specific traffic and total PCAP content. IKE bytes were calculated by aggregating frame lengths across all packets within each exchange type category, while total PCAP bytes encompass all captured network traffic including non-IKE communications. This measurement approach enables precise quantification of protocol overhead introduced by different cryptographic approaches, particularly the additional \texttt{IKE\_INTERMEDIATE} exchanges required for hybrid QKD-PQC key establishment, while also providing context on the proportion of IKE traffic relative to overall network activity. The total protocol overhead is calculated as the sum of bytes transmitted across all IKE exchange types during the complete handshake sequence, providing a metric for comparing the network efficiency of traditional cryptographic proposals against QKD-enhanced alternatives.

\subsubsection{Timing Measurement Definitions}

Our performance analysis distinguishes between fundamental timing metrics to isolate network communication overhead from additional computational processing costs, defined through the following measurable quantities:

\paragraph{Network Time Measurement ($t_{\text{net}}$)} Network latency measurements were conducted from the initiator perspective (Alice) to capture end-to-end handshake performance including both network propagation delays and responder processing overhead. Packet timestamps were extracted from PCAP captures performed on Alice's network interface, providing round-trip timing measurements that encompass the complete request-response cycle. To isolate the core key establishment phase from authentication overhead, latency was measured from the first \texttt{IKE\_SA\_INIT} packet to the last message preceding the first \texttt{IKE\_AUTH} packet within each iteration. This measurement window specifically captures the Diffie-Hellman/KEM exchange (\texttt{IKE\_SA\_INIT}) and any additional key material exchanges (\texttt{IKE\_INTERMEDIATE} for QKD/PQC proposals) while excluding identity verification and authentication phases. Importantly, this network time measurement includes the time required for blocking HTTPS calls to QKD nodes during key retrieval operations, as these calls occur synchronously within the handshake protocol flow.

\paragraph{Plugin Execution Time Measurement ($t_{\text{plugin}}$)} Plugin execution time represents the total wall-clock time during which our strongSwan plugins are actively engaged in cryptographic operations. This metric is measured by instrumenting our QKD and QKD-KEM plugins to record timestamps at the beginning of the first cryptographic operation ($t_{\text{create}}$) and at the completion of the final operation ($t_{\text{destroy}}$) within each handshake iteration:
\begin{equation}
t_{\text{plugin}} = t_{\text{destroy}} - t_{\text{create}}
\end{equation}

For sequential hybrid proposals involving multiple \texttt{IKE\_INTERMEDIATE} exchanges, the plugin execution time spans from the first plugin creation event to the last plugin destruction event across all key establishment phases, effectively capturing the cumulative time during which quantum or hybrid cryptographic operations are performed. Mathematically, for a proposal involving $N$ individual cryptographic exchanges indexed by $i = 1, 2, \ldots, N$, each with creation timestamp $t_{\text{create},i}$ and destruction timestamp $t_{\text{destroy},i}$, the total plugin execution time is computed as:
\begin{equation}
t_{\text{plugin}} = \max_{i=1,\ldots,N} \{t_{\text{destroy},i}\} - \min_{i=1,\ldots,N} \{t_{\text{create},i}\}
\end{equation}

\paragraph{Plugin Processing Additional Overhead ($\Delta t_{\text{overhead}}$)} The plugin processing additional overhead isolates the computational cost attributable to our cryptographic plugin operations beyond what is already accounted for in the network communication time
\begin{equation}
\Delta t_{\text{overhead}} = t_{\text{plugin}} - t_{\text{net}}
\end{equation}

This metric identifies additional computational processing time that occurs outside of network communication periods (as accounted by traffic capture). However, QKD operations can appear in both timing categories depending on their execution context: QKD key retrieval operations that occur simultaneously with strongSwan's \texttt{set\_public\_key} primitive (performing KEM encapsulation) within the \texttt{IKE\_SA\_INIT} exchange are captured within $t_{\text{net}}$, while QKD operations that occur during strongSwan's \texttt{get\_public\_key} primitive (performing KEM key generation) or strongSwan's \texttt{set\_public\_key} primitive (performing KEM decapsulation) in the last proposal of sequential exchanges are captured in $\Delta t_{\text{overhead}}$. The specific timing allocation also depends on whether client-initiated or server-initiated QKD flows are used, as these affect when QKD key retrieval occurs relative to network message exchanges and their corresponding strongSwan key exchange primitives. 

A positive value $\Delta t_{\text{overhead}} > 0$ indicates additional computational processing beyond basic network communication, while values approaching zero ($\Delta t_{\text{overhead}} \approx 0$) suggest that plugin operations occur predominantly during network wait periods or involve primarily network-bound operations.

\paragraph{Statistical Analysis and Error Propagation} Since our performance evaluation involves multiple experimental iterations for each cryptographic proposal, we characterize measurement uncertainty through standard deviations of the directly measured quantities $t_{\text{net}}$ and $t_{\text{plugin}}$. However, these measurements are not independent: the plugin execution time $t_{\text{plugin}}$ encompasses the network communication period $t_{\text{net}}$, establishing the relationship:
\begin{equation}
t_{\text{plugin}} = t_{\text{net}} + \Delta t_{\text{overhead}}
\end{equation}

For the derived quantity $\Delta t_{\text{overhead}} = t_{\text{plugin}} - t_{\text{net}}$, the proper uncertainty estimation must account for the correlation between these measurements. In the limiting case where $t_{\text{net}}$ and $t_{\text{plugin}}$ are perfectly correlated (i.e., when $\Delta t_{\text{overhead}} \approx 0$), the uncertainties would partially cancel. However, since we lack direct measurements of the correlation coefficient and the overhead represents genuinely independent computational processes occurring outside the network communication periods, we adopt a conservative approach by treating the measurements as independent for uncertainty estimation:
\begin{equation}
\sigma_{\Delta t_{\text{overhead}}} = \sqrt{\sigma_{t_{\text{plugin}}}^2 + \sigma_{t_{\text{net}}}^2}
\end{equation}

This approach provides an upper bound on the true uncertainty, ensuring that our reported error bars are conservative estimates that account for the maximum possible propagation of measurement uncertainties in the derived overhead metric.

Complete implementation details, including configuration parameters and source code, are available in our public repository \cite{qkd-ipsec-docker}, enabling other researchers to reproduce our experiments. The \texttt{README.md} contains detailed instructions. The IKEv2 exchanges can be monitored on UDP ports 500 and 4500 using Wireshark.

It is important to note that our performance measurements focus on qualitative comparisons between different key establishment methodologies rather than absolute latency values. The absolute timing results are highly dependent on implementation details, network topology, and QKD infrastructure characteristics such as KME proximity, key buffer availability, and backend response times. Our primary objective is to analyze the relative performance differences between sequential and parallel hybridization approaches, the impact of network conditions on different cryptographic proposals, and the protocol overhead introduced by fragmentation and additional \texttt{IKE\_INTERMEDIATE} exchanges.

\section{Results}\label{sec:results}

To understand the fundamental impact of post-quantum cryptography on IPsec protocol efficiency, we first examine the packet-level overhead introduced by different cryptographic algorithms. Table~\ref{tab:fragmentation_analysis} presents the observed \texttt{IKE\_SA\_INIT} packet sizes captured via Wireshark analysis across our tested cryptographic proposals. The measurements reveal the bandwidth overhead differences between classical, post-quantum, and QKD hybrids. Classical algorithms like ECC P-256 require minimal space (334-367 bytes), while QKD maintains the smallest footprint (270-319 bytes) due to its identifier-based approach. In contrast, post-quantum algorithms impose significant overhead: FrodoKEM variants require extensive fragmentation with the largest algorithms (FrodoA5/FrodoS5) generating over 22KB of data across 15 fragments, while HQC algorithms show asymmetric behavior with responder messages requiring substantially more fragments than initiator messages. The observed fragment sizes of 1,514 bytes align with our strongSwan configuration of 1,472-byte fragments plus IP/UDP headers, confirming that our buffer and fragmentation settings successfully accommodate the largest post-quantum payloads without processing failures. This data provides the foundation for understanding the subsequent performance and timing measurements.

\begin{table*}[htbp]
\centering
\caption{\texttt{IKE\_SA\_INIT} packet fragmentation analysis across cryptographic algorithms}
\label{tab:fragmentation_analysis}
\resizebox{\textwidth}{!}{%
\begin{tabular}{|l|c|c|c|c|c|c|c|c|c|c|c|c|}
\hline
\multirow{2}{*}{\textbf{Algorithm}} & \multicolumn{3}{c|}{\textbf{Initiator}} & \multicolumn{3}{c|}{\textbf{Responder}} & \multicolumn{2}{c|}{\textbf{Crypto Payload}} & \multicolumn{2}{c|}{\textbf{Total Size}} & \multicolumn{2}{c|}{\textbf{IPsec Overhead}} \\
\cline{2-13}
& \textbf{Frags} & \textbf{Last} & \textbf{Avail} & \textbf{Frags} & \textbf{Last} & \textbf{Avail} & \textbf{PK} & \textbf{CT} & \textbf{Init} & \textbf{Resp} & \textbf{Init} & \textbf{Resp} \\
& & \textbf{(B)} & \textbf{(B)} & & \textbf{(B)} & \textbf{(B)} & \textbf{(B)} & \textbf{(B)} & \textbf{(B)} & \textbf{(B)} & \textbf{(B)} & \textbf{(B)} \\
\hline
X25519 & 1 & 302 & 1212 & 1 & 335 & 1179 & 32 & 32 & 302 & 335 & 270 & 303 \\
\hline
ECP256 & 1 & 334 & 1180 & 1 & 367 & 1147 & 64 & 64 & 334 & 367 & 270 & 303 \\
\hline
bike1 & 2 & 331 & 1183 & 2 & 396 & 1118 & 1541 & 1573 & 1845 & 1910 & 304 & 337 \\
\hline
bike3 & 3 & 393 & 1121 & 3 & 458 & 1056 & 3083 & 3115 & 3421 & 3486 & 338 & 371 \\
\hline
frodoa1 & 7 & 1006 & 508 & 7 & 1143 & 371 & 9616 & 9720 & 10090 & 10227 & 474 & 507 \\
\hline
frodoa3 & 11 & 1102 & 412 & 11 & 1247 & 267 & 15632 & 15744 & 16242 & 16387 & 610 & 643 \\
\hline
frodoa5 & 15 & 1070 & 444 & 15 & 1215 & 299 & 21520 & 21632 & 22266 & 22411 & 746 & 779 \\
\hline
frodos1 & 7 & 1006 & 508 & 7 & 1143 & 371 & 9616 & 9720 & 10090 & 10227 & 474 & 507 \\
\hline
frodos3 & 11 & 1102 & 412 & 11 & 1247 & 267 & 15632 & 15744 & 16242 & 16387 & 610 & 643 \\
\hline
frodos5 & 15 & 1070 & 444 & 15 & 1215 & 299 & 21520 & 21632 & 22266 & 22411 & 746 & 779 \\
\hline
hqc1 & 2 & 1039 & 475 & 4 & 296 & 1218 & 2249 & 4433 & 2553 & 4838 & 304 & 405 \\
\hline
hqc3 & 4 & 352 & 1162 & 7 & 401 & 1113 & 4522 & 8978 & 4894 & 9485 & 372 & 507 \\
\hline
hqc5 & 6 & 115 & 1399 & 10 & 1404 & 110 & 7245 & 14421 & 7685 & 15030 & 440 & 609 \\
\hline
kyber1 & 1 & 1070 & 444 & 1 & 1071 & 443 & 800 & 768 & 1070 & 1071 & 270 & 303 \\
\hline
kyber3 & 1 & 1454 & 60 & 1 & 1391 & 123 & 1184 & 1088 & 1454 & 1391 & 270 & 303 \\
\hline
kyber5 & 2 & 358 & 1156 & 2 & 391 & 1123 & 1568 & 1568 & 1872 & 1905 & 304 & 337 \\
\hline
qkd & 1 & 270 & 1244 & 1 & 319 & 1195 & 0 & 16 & 270 & 319 & 270 & 303 \\
\hline
\end{tabular}%
}
\captionsetup{font=footnotesize}
\caption*{\textbf{Notes:} Measurements captured via Wireshark showing complete Ethernet frames. Frags = fragment count; Last = last fragment size; Avail = available space in last fragment (1514 - Last); PK = public key size; CT = ciphertext size; Init/Resp = initiator/responder total size. IPsec overhead represents protocol headers and metadata: 34 bytes per fragment plus 236 bytes (initiator) and 269 bytes (responder) for IKE protocol structure. Total sizes calculated as $(\text{fragments}-1) \times 1514 + \text{last fragment size}$. QKD uses 16-byte key identifiers instead of traditional public keys, achieving minimal bandwidth overhead.}
\end{table*}

To evaluate the impact of network conditions on different cryptographic approaches, we selected a representative subset of proposals that isolate key performance characteristics. Our selection encompasses scenarios with minimal fragmentation overhead, such as Kyber512 which transmits within single \texttt{IKE\_SA\_INIT} messages, and scenarios requiring fragmentation, such as Kyber1024 where both initiator and responder messages fragment into two packets. We include both parallel hybrid implementations using our QKD-KEM provider (qkd\_kyber1, qkd\_kyber5) and sequential hybrid implementations utilizing RFC 9370's \texttt{IKE\_INTERMEDIATE} exchanges (qkd-ke1\_kyber1, kyber1-ke1\_qkd, qkd-ke1\_kyber5, kyber5-ke1\_qkd). The sequential hybrids are tested in both key establishment orders (i.e. QKD followed by post-quantum (qkd-ke1\_kyber) and post-quantum followed by QKD (kyber-ke1\_qkd)) to assess the impact of key exchange sequencing on overall handshake performance. Additionally, we include pure QKD and classical reference algorithms (ECC P-256, X25519) alongside a traditional sequential hybrid (x25519-ke1\_kyber1) to provide baseline comparisons across cryptographic paradigms.

\begin{figure*}[tbp]
    \centering
    \includegraphics[width=0.9\textwidth]{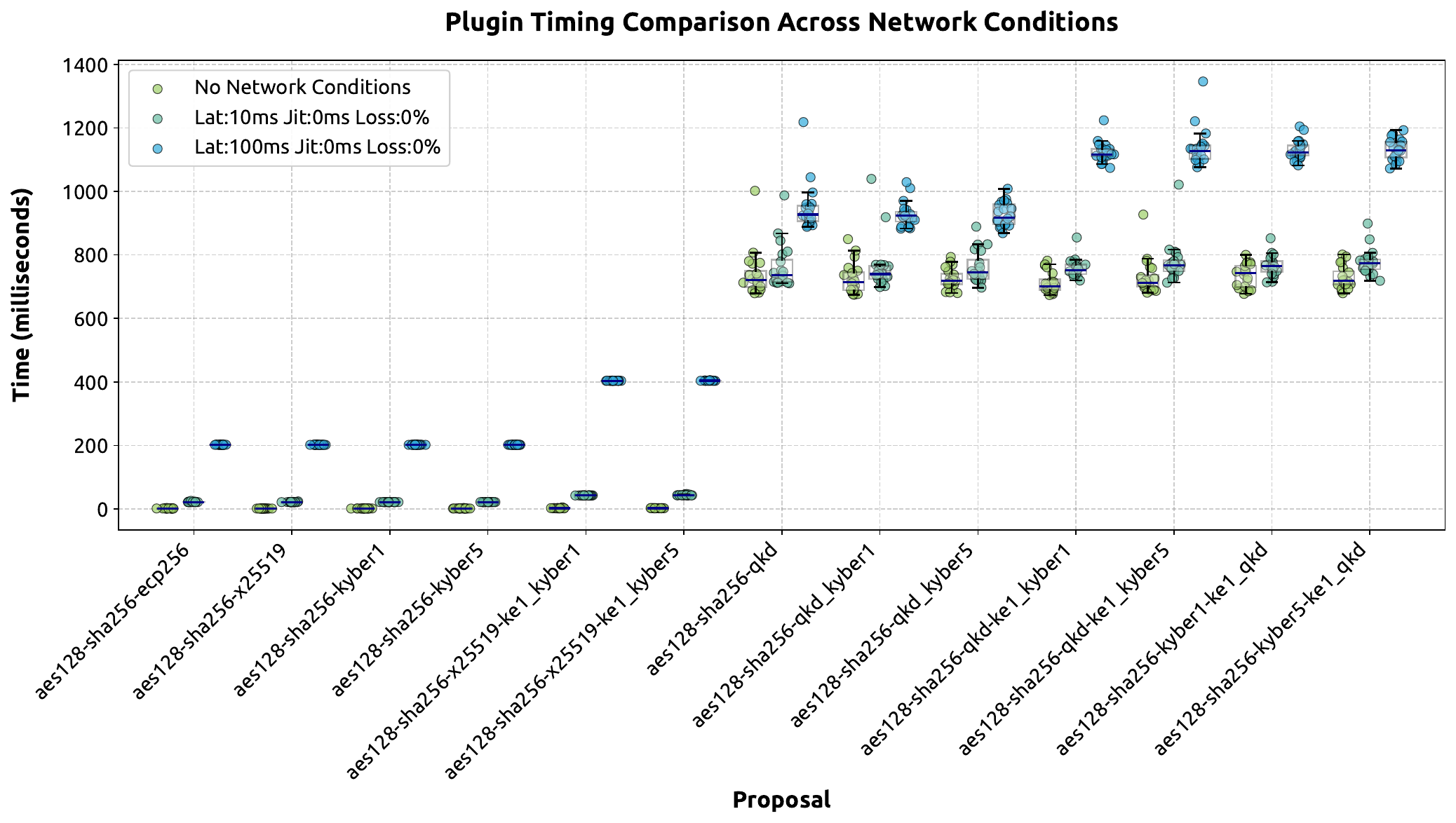}
    \caption{Average time per iteration for the different proposals with different values for latency (no network conditions, 10ms and 100ms)}
    \label{fig:latencies_delay}
\end{figure*}

Figure \ref{fig:latencies_delay} shows the plugin execution time measurements ($t_{\text{plugin}}$) with several network delays affecting the communication between Alice and Bob (but not their respective communication with the QKD nodes).

\begin{figure}[tp]
    \centering
    \includegraphics[width=1\columnwidth]{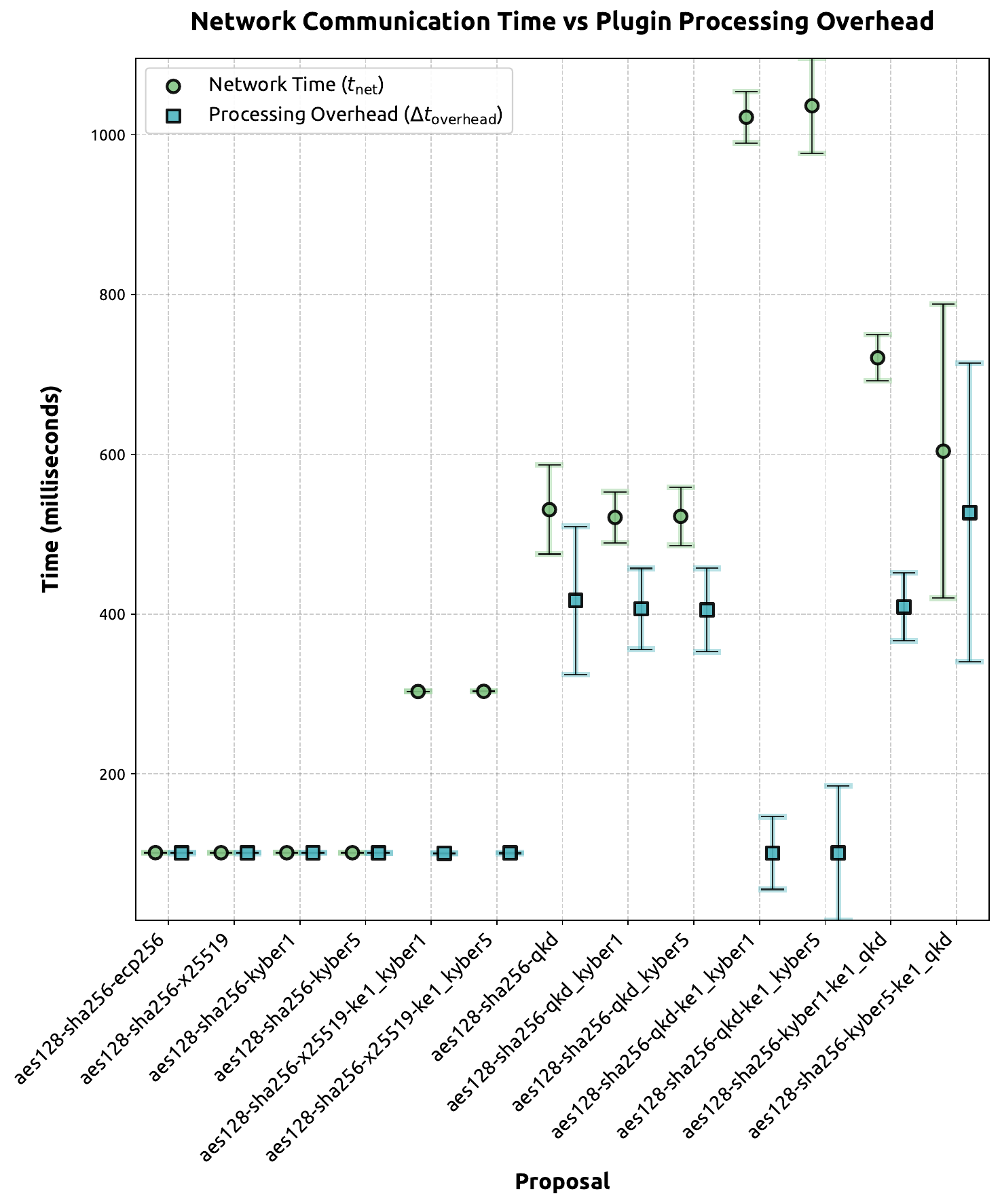}
    \caption{Network communication time ($t_{\text{net}}$) versus plugin processing additional overhead ($\Delta t_{\text{overhead}}$) for different cryptographic proposals under 100ms artificial network delay. Network time (green circles) captures the complete IKE handshake duration from Alice's perspective, while processing overhead (blue squares) represents additional computational costs. The proposal ordering affects network time due to ETSI API differences: \texttt{qkd\_ke1\_kyber} uses client-initiated QKD flow while \texttt{kyber1\_ke1\_qkd} uses server-initiated flow with additional \texttt{GET\_KEY\_WITH\_IDS()} latency.}
    \label{fig:network_vs_processing}
\end{figure}

\begin{figure}[tp]
    \centering
    \includegraphics[width=1\columnwidth]{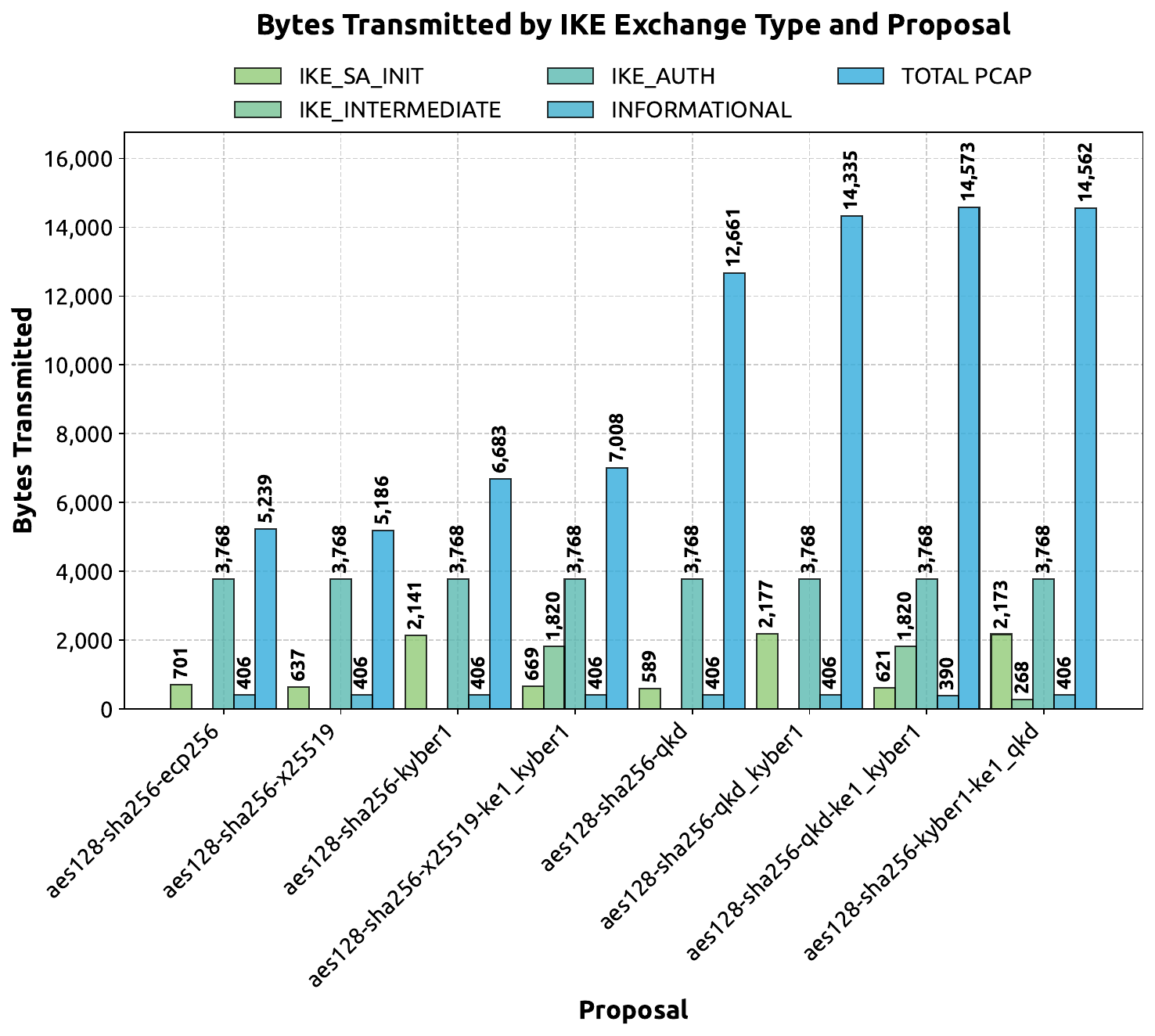}
    \caption{Bytes per message type}
    \label{fig:ike_bytes}
\end{figure}

To further decompose the performance characteristics, Figure \ref{fig:network_vs_processing} provides a breakdown of network communication time ($t_{\text{net}}$) versus plugin processing additional overhead ($\Delta t_{\text{overhead}}$) across different cryptographic proposals under 100ms artificial network delay. This decomposition isolates the contribution of network latency from computational overhead, revealing how different hybridization strategies impact protocol efficiency.

The bytes per message type are shown in Figure \ref{fig:ike_bytes}, which quantifies the additional bandwidth consumption introduced by \texttt{IKE\_INTERMEDIATE} exchanges in sequential hybridization approaches. The results show increased total bytes transmitted compared to parallel hybridization methods.

\section{Discussion}
\label{sec:discussion}

While previous research has proposed various approaches for integrating QKD with IPsec protocols, the literature lacks comprehensive comparative analyses between sequential and parallel hybrid key establishment strategies under realistic network conditions. Prior work such as the AQUA protocol~\cite{nagayama-ipsecme-ipsec-with-qkd-01}, commercial implementations by FortiOS~\cite{fortinet2024qkd} and Juniper~\cite{juniper2025qkd}, and recent field trials~\cite{sibson2024field, alia2024100} have primarily focused on functional integration without systematic evaluation of protocol performance trade-offs. Specifically, existing studies have not provided detailed analysis of how different hybridization approaches perform under varying network conditions including latency, packet loss, and fragmentation scenarios, nor have they quantified the bandwidth overhead differences between RFC 9370's sequential \texttt{IKE\_INTERMEDIATE} exchanges and parallel hybrid implementations. Our work addresses this gap by providing the first comprehensive performance comparison of QKD hybrid key establishment strategies, revealing that parallel approaches can achieve significant performance advantages over standardized sequential methods, particularly in high-latency environments where each additional round-trip compounds network delays.

An important design consideration for hybrid QKD key establishment concerns the asymmetric payload sizes between KEM public keys and ciphertexts, which significantly impact fragmentation overhead depending on flow initiation choice. As evident in Table~\ref{tab:fragmentation_analysis}, post-quantum algorithms exhibit varying asymmetry: HQC-128 shows 2,249-byte public keys (2 fragments) versus 4,433-byte ciphertexts (4 fragments), while ML-KEM algorithms like Kyber-512 demonstrate balanced 800-byte public keys and 768-byte ciphertexts (both single fragments). For algorithms where ciphertexts exceed public key sizes, server-initiated flows could reduce fragmentation by placing larger ciphertext payloads alongside small QKD identifiers (16 bytes) in responder messages. However, server-initiated flows require ETSI GS QKD 014's stateless interface, limiting optimization flexibility. Future hybrid key establishment protocols could benefit from adaptive flow selection based on algorithm characteristics to minimize fragmentation overhead while respecting QKD backend API constraints.

Our experimental results highlight important considerations regarding the RFC 9242 design strategy for quantum-resistant cryptography deployment. RFC 9242 introduced the \texttt{\texttt{IKE\_INTERMEDIATE}} exchange to address fragmentation-related issues (including firewall fragment dropping, stateful inspection failures, and reduced network reliability) by providing application-level fragmentation capabilities. Our latency measurements demonstrate that this fragmentation avoidance strategy introduces significant performance penalties in high-latency environments, where each additional round-trip incurs the full network delay. In our controlled network environment, we observed IP fragmentation occurring only for ML-KEM-1024 (Kyber Level 5) combinations, where payload sizes of 1584 bytes exceed the standard 1500-byte Ethernet Maximum Transmission Unit (MTU). This IP-level fragmentation showed no measurable latency impact, as fragments are transmitted simultaneously and reassembled at the network layer, preserving the single logical round-trip nature of the key exchange; but in realistic networks, this may not be the case.

The sequential \texttt{IKE\_INTERMEDIATE} approach mandated by RFC 9370 for hybrid key exchanges creates multiplicative latency penalties that significantly impact handshake performance. For QKD+ML-KEM hybrid combinations that naturally fit within MTU limits (such as QKD+ML-KEM-512 (816 bytes) and QKD+ML-KEM-768 (1200 bytes)) the parallel hybridization approach eliminates both fragmentation concerns and protocol overhead simultaneously. This analysis suggests that protocol design should prioritize payload size estimation: parallel hybridization should be preferred when combined payloads remain within Path MTU limits, as it avoids the latency penalties associated with sequential exchanges and the extra bytes being transmitted (due to packet headers) while maintaining compatibility with existing network infrastructure. Sequential approaches with \texttt{IKE\_INTERMEDIATE} remain valuable for scenarios where fragmentation is unavoidable or where network reliability concerns regarding fragment handling outweigh latency optimization requirements.

Our experimental results demonstrate that the parallel hybridization approach implemented through the QKD-KEM plugin consistently outperforms the sequential hybridization method using RFC 9370's \texttt{IKE\_INTERMEDIATE} exchanges. The superior performance of the parallel approach can be attributed to reduced protocol overhead, as it eliminates the need for additional round trips required by the sequential \texttt{IKE\_INTERMEDIATE} exchanges.

Network latency testing significantly amplifies the performance differences between sequential and parallel hybridization approaches. As shown in Figure \ref{fig:latencies_delay}, under 500ms network latency conditions, the performance gap widens considerably compared to baseline measurements.
This divergence stems from architectural differences between the two approaches. The sequential hybridization method requires additional network round-trips through \texttt{IKE\_INTERMEDIATE} exchanges, with each round-trip incurring the full latency penalty. In contrast, the parallel approach implemented through our QKD-KEM plugin maintains relatively stable performance under high-latency conditions by preserving the original IKEv2 message exchange pattern without additional round-trips.
These findings have significant implications for deployments involving satellite links, international connections, or mobile networks where latency can be substantial. Our tests with packet loss further revealed that the sequential approach is more vulnerable to connection failures in lossy networks due to its increased packet exchange. This reinforces the advantage of minimizing protocol round-trips in challenging network environments, particularly for security protocols where handshake reliability is critical.

Moreover, bandwidth analysis (Figure~\ref{fig:ike_bytes}) shows how the hybrid exchange via \texttt{IKE\_INTERMEDIATE} requires more total bytes.

In this test, this applies to our QKD hybrids, but this should apply, for example, to traditional + PQC hybrids as well, as the extra bytes come from the extra packets required (that include headers and so on, not strictly necessary for the exchange of key material during the key establishment).

Figure \ref{fig:network_vs_processing} shows the decomposition of total handshake time into network latency $t_{\text{net}}$ and plugin processing additional overhead $\Delta t_{\text{overhead}}$ components across different cryptographic proposals under 100ms artificial network delay conditions. The sequential QKD hybrid proposals exhibit significantly higher $t_{\text{net}}$ compared to traditional and parallel approaches due to the additional \texttt{IKE\_INTERMEDIATE} round-trips mandated by RFC 9370, with each round-trip incurring the full 100ms latency penalty. The $\Delta t_{\text{overhead}}$ component remains relatively constant across QKD-based proposals, as the same cryptographic operations are performed regardless of the hybridization method. The elevated $t_{\text{net}}$ observed in QKD proposals compared to traditional cryptography also reflects the additional round-trip required for Bob to retrieve quantum keys from the QKD node during the handshake process. This analysis confirms that performance differences between sequential and parallel hybridization approaches are primarily attributable to network-level protocol overhead rather than computational complexity, reinforcing the architectural advantages of the parallel QKD-KEM approach in latency-sensitive environments.

The network timing measurements $t_{\text{net}}$ were captured from Alice's (initiator) perspective using packet capture analysis, measuring the complete IKE handshake duration from the first \texttt{IKE\_SA\_INIT} packet through all key establishment exchanges, including \texttt{IKE\_INTERMEDIATE} rounds and any retransmissions. The ordering of proposals in hybrid combinations significantly affects $t_{\text{net}}$ due to the ETSI API operational differences. In the \texttt{qkd-ke1\_kyber} sequence, the QKD key retrieval occurs first using the client-initiated flow, followed by the Kyber exchange in an \texttt{IKE\_INTERMEDIATE} round. Conversely, in the \texttt{kyber1-ke1\_qkd} sequence, Kyber is exchanged first, and QKD key retrieval happens in the second \texttt{IKE\_INTERMEDIATE} using the server-initiated flow, where Bob retrieves the QKD key and Alice subsequently calls \texttt{GET\_KEY\_WITH\_IDS()}. This server-initiated QKD operation introduces additional latency, explaining why \texttt{kyber1-ke1\_qkd} exhibits higher $t_{\text{net}}$ than \texttt{qkd-ke1\_kyber}.

Notably, pure QKD and the parallel hybrid \texttt{qkd\_kyber1} exhibit similar $t_{\text{net}}$ because both require the same number of network round-trips, while \texttt{kyber1-ke1\_qkd} requires two additional \texttt{IKE\_INTERMEDIATE} exchanging in approximately 200ms additional latency under the 100ms artificial delay condition. This 200ms difference directly corresponds to the two extra trips required by the sequential approach, confirming that each \texttt{IKE\_INTERMEDIATE} exchange incurs the full network latency penalty. Similar $t_{\text{net}}$ increases are observed when comparing \texttt{x25519-ke1\_kyber1} (sequential classical-PQC hybrid) with non-hybrid algorithms, demonstrating that the latency penalty is inherent to the sequential hybridization strategy rather than specific to QKD integration.

While our experimental results provide concrete timing measurements, it is crucial to emphasize that the absolute latency values should not be considered definitive performance benchmarks for QKD-IPsec deployments. The observed latencies are heavily influenced by our specific testbed configuration, including the physical separation between Docker containers and QKD nodes, network routing through our gateway architecture, and the particular ETSI API implementation characteristics. In production deployments with optimized QKD infrastructure (such as co-located KMEs with IPsec endpoints, pre-buffered key pools, and optimized classical communication channels between applications and KMEs) absolute latencies for QKD operations could be reduced significantly. The fundamental value of our analysis lies in the comparative assessment of different key establishment strategies: demonstrating that parallel hybrid approaches consistently outperform sequential methods regardless of absolute timing, quantifying the multiplicative impact of additional round-trips in high-latency environments, and revealing how fragmentation overhead varies across different cryptographic algorithms. These qualitative insights remain valid across diverse deployment scenarios and provide practical guidance for selecting appropriate hybridization strategies based on network characteristics and performance requirements.

The recent NIST standardization of ML-KEM (Module-Lattice-Based Key Encapsulation Mechanism), formerly known as CRYSTALS-Kyber, as FIPS 203~\cite{fips203}, makes our QKD+ML-KEM hybrid implementations particularly significant for practical deployments. Our results demonstrate that QKD+ML-KEM-512 (816 bytes total payload) and QKD+ML-KEM-768 (1200 bytes total payload) combinations fit comfortably within standard 1500-byte MTU limits without requiring fragmentation, while providing both NIST-approved post-quantum security and information-theoretic security from QKD. This compatibility is crucial because ML-KEM represents the primary standardized post-quantum KEM algorithm that organizations will deploy to meet quantum-resistant cryptography requirements. Unlike other post-quantum algorithms such as HQC variants that require extensive fragmentation (up to 10 fragments for HQC-256), ML-KEM's balanced public key and ciphertext sizes make it an ideal candidate for QKD hybridization. The combination of standardized post-quantum cryptography with quantum key distribution through our parallel hybrid approach provides a practical defense-in-depth strategy that organizations can implement today using established standards, avoiding both the fragmentation overhead of larger post-quantum algorithms and the protocol latency penalties of sequential approaches.

\section{Conclusions and Future Work}\label{sec:conclusions}

We have demonstrated a novel quantum-resistant security framework for IKEv2/IPsec with broader applicability to secure communication protocols like TLS. Building upon the foundational concepts introduced in the AQUA protocol by Nagayama and Van Meter~\cite{nagayama-ipsecme-ipsec-with-qkd-01}, which pioneered the use of QKD key identifiers as replacements for traditional Diffie-Hellman exchanges, our work advances this paradigm through two distinct architectural contributions. First, we have developed a sequential hybridization method that leverages the RFC 9370 framework to systematically combine QKD-derived material with both classical and post-quantum cryptographic elements. Second, we have created an integrated QKD-KEM abstraction that unifies quantum key management within standardized KEM interfaces. These complementary approaches embody contrasting design philosophies: sequential hybridization preserves modular separation of cryptographic components, while the unified QKD-KEM abstraction achieves streamlined integration with minimized protocol complexity. Through our strongSwan implementation framework, we have successfully demonstrated compatibility across both ETSI GS QKD 004~\cite{etsi2020qkd} and ETSI GS QKD 014~\cite{etsi2020qkd014} specifications, thereby resolving practical interoperability challenges encountered when deploying commercial QKD infrastructure.

This finding has important implications for IPsec implementations requiring both quantum resistance and optimized handshake performance, particularly in latency-sensitive applications that can still benefit from the information-theoretic security guarantees of QKD. A limitation of our integration is the lack of explicit proof-of-possession, unlike Pr\'evost et al.~\cite{prevost2025etsi}, potentially leaving it vulnerable to replay attacks without active key verification or mutual possession proof. Future work includes implementing ETSI 004 index management for session reuse across rekeying operations in both pure QKD and hybrid approaches, which would further reduce overhead for long-lived IKE SAs by avoiding repeated KSID transmissions; incorporating explicit possession proofs to enhance security; and using RFC 9242’s IKE\_INTERMEDIATE for the implementation of a proof-of-possession mechanism.

\section*{Acknowledgment}
This work was supported by the Spanish Government under the following grants funded by MICIU/AEI/10.13039/501100011033: (i) ``QUantum-based ReSistant Architectures and Techniques (QURSA)" TED-2021-130369B-C32 and by the "European Union NextGenerationEU/PRTR"; (ii) ``DIstributed Smart Communications with Verifiable EneRgy-optimal Yields (DISCOVERY)" PID2023-148716OB-C32. In addition, it was partially supported by the I-Shaper Strategic Project (PRTR-INCIBE-2023/00623/001), funded by the "European Union NextGenerationEU/PRTR" due to the collaboration agreement signed between the Instituto Nacional de Ciberseguridad (INCIBE) and the UC3M; this initiative is being carried out within the framework of the Recovery, Transformation and Resilience Plan funds, funded by the European Union (Next Generation). It was also supported by the Comunidad de Madrid under the grant ``RAMONES-CM" TEC-2024/COM-504.

\printbibliography
\end{document}